\documentclass{article}
\usepackage{spconf,amsmath,graphicx,booktabs,color,makecell,stfloats,amsfonts,amssymb,enumitem, tabularx}
\usepackage[skip=4pt]{caption}

\title{Semantic Proximity Alignment: Towards Human Perception-consistent \\ Audio Tagging by Aligning with Label Text Description}
\name{Wuyang Liu$^2$, Yanzhen Ren$^{1,2,\ast}$ \thanks{$^{\ast}$Corresponding author.}\thanks{This work is supported by the Natural Science Foundation of China (NSFC) under the grant NO. 62172306, Hubei Province Technological Innovation Major Project (NO. 2021BAA034, 2020BAB018)
}}
\address{$^1$Key Laboratory of Aerospace Information Security and Trusted Computing, Ministry of Education, \\ $^2$School of Cyber Science and Engineering, Wuhan University}
\begin{document}
%
\maketitle
\begin{abstract}
Most audio tagging models are trained with one-hot labels as supervised information. However, one-hot labels treat all sound events equally, ignoring the semantic hierarchy and proximity relationships between sound events. In contrast, the event descriptions contains richer information, describing the distance between different sound events with semantic proximity.
In this paper, we explore the impact of training audio tagging models with auxiliary text descriptions of sound events. By aligning the audio features with the text features of corresponding labels, we inject the hierarchy and proximity information of sound events into audio encoders, improving the performance while making the prediction more consistent with human perception. We refer to this approach as Semantic Proximity Alignment (SPA).
We use Ontology-aware mean Average Precision (OmAP) as the main evaluation metric for the models. OmAP reweights the false positives based on Audioset ontology distance and is more consistent with human perception compared to mAP. Experimental results show that the audio tagging models trained with SPA achieve higher OmAP compared to models trained with one-hot labels solely (+1.8 OmAP). Human evaluations also demonstrate that the predictions of SPA models are more consistent with human perception.

\end{abstract}
\begin{keywords}
Audio Classification, Sound Event Detection, Multi-modality, Audio-text Pretraining, Transformer
\end{keywords}
\section{Introduction}
\label{sec:intro}

Audio tagging refers to tagging an audio recording with one or more audio events. It has broad applications such as audio surveillance \cite{park2021pretrained}, noise control \cite{bello2019sonyc}, etc.
The largest publicly available audio tagging dataset, Audioset \cite{gemmeke2017audio}, is commonly used to train audio tagging models. As a multi-label classification task, state-of-the-art models on Audioset like \cite{huang2022mavil, xu2022masked, pmlr-v202-chen23ag} are trained under a classical supervised setting that adopts one-hot labels as targets, binary cross entropy (BCE) as loss function and mean average precision (mAP) as evaluation metric.

The problems of this setup, as discussed in \cite{sun2020ontology, aironi2022graph,  liu2023ontology}, are twofold. First, Audioset is known to have many incorrect or missing labels, which means the models is not always properly guided even when its prediction is acceptable. Second, the inter-class relationship between audio classes are not naturally exclusive, which means that audio classes can be similar to (e.g., \textit{Laughter} and \textit{Giggle}) or contain (e.g., \textit{Speech} and \textit{Male Speech}) some classes. 
To address these problems, graph convolutional network (GCN) is adopted in \cite{sun2020ontology, aironi2022graph} to utilize Audioset ontology infomation. Ontology-aware binary cross entropy (OBCE) and ontology-aware mean average precision (OmAP) are proposed in \cite{liu2023ontology}, which reweights the false positives based on the ontology distance between the prediction and the target label.

However, despite the attempts to incorporate the hierarchical information from the Audioset ontology, the supervision information in \cite{sun2020ontology, aironi2022graph,  liu2023ontology} is still provided by one-hot labels solely, which treats all sound classes equally and ignore the semantic hierarchy and proximity relationships between sound events. In the last two years, many approaches have explored the combination of audio and text modalities to train audio classification models \cite{xie2021zero, elizalde2023clap, wu2022wav2clip, guzhov2022audioclip}. These methods emphasize the flexibility of natural language supervision and evaluate the performance on zero-shot classification tasks as the original CLIP \cite{radford2021learning} proposed. Other methods \cite{islam2019soundsemantics, shi2020few, zhang2022wikitag} have focused on the impact of rich text sources in audio tagging. They distill auxiliary knowledge from text into the audio encoders to achieve few-shot classification. The aforementioned text-audio multimodal works did not evaluate the benefit of natural language supervision to classical supervised training setup.

In this paper, we propose Semantic Proximity Alignment (SPA) to use natural language as auxiliary supervision information when training audio tagging models. Instead of directly mapping audio features to one-hot labels, we first extract the text features of pre-constructed language description of each audio class with a pretrained language model, and then align the audio feature of each clip with corresponding text features of its positive labels using proposed SPA loss. This way, we inject the semantic information present in natural language into the audio model, allowing the model to learn hierarchy and proximity relationships between sound events. Experimental results demonstrate that audio models trained with SPA achieve higher OmAP than those without. Further human opinion experiments also confirm that the predictions of these models better align with human perception.

\section{Analysis of label representation}
\label{sec:2}

\begin{figure}[ht]
	\centering
	\includegraphics[width=8.5cm]{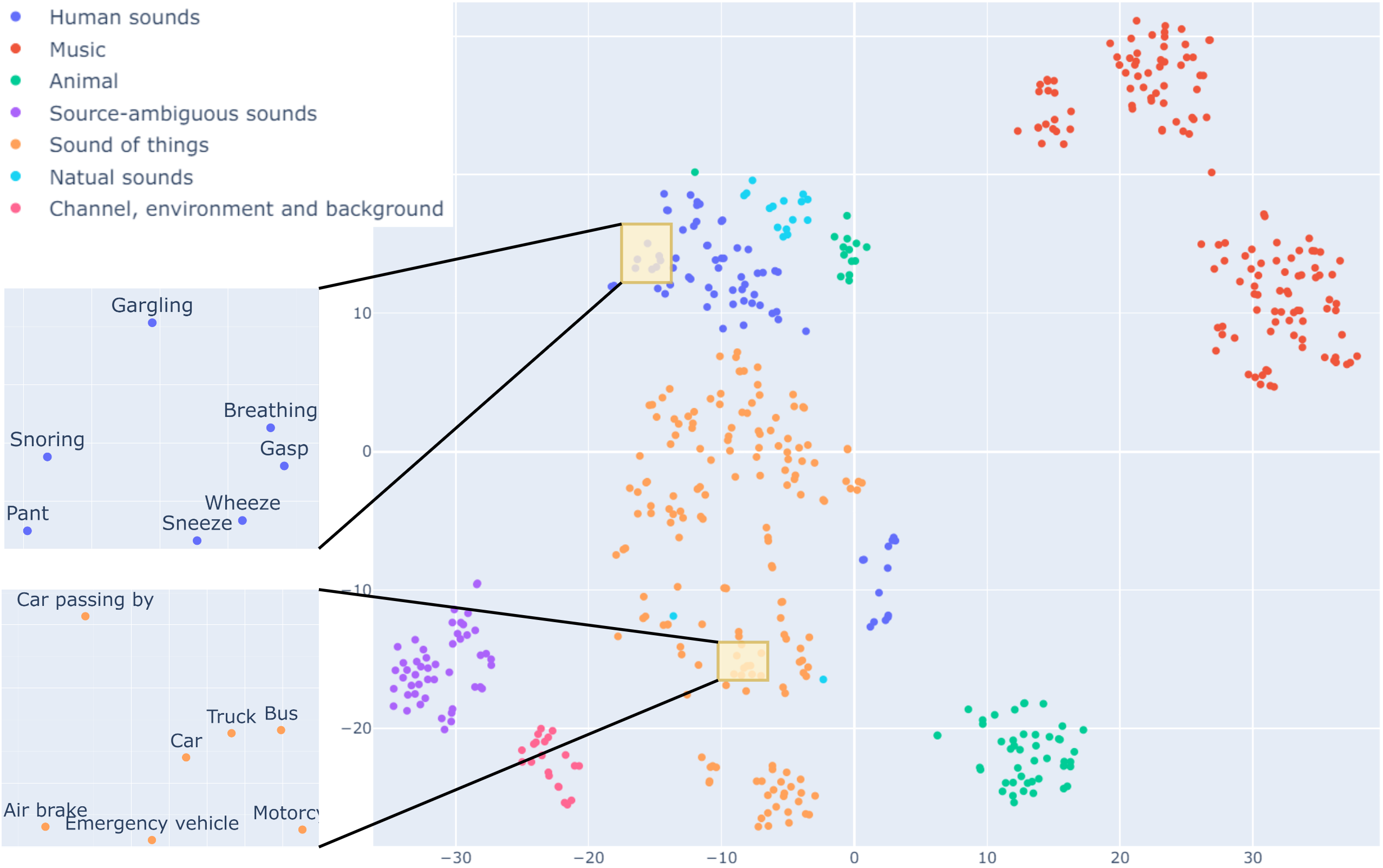}
	\caption{t-SNE visualization of the text embeddings of the language description of 527 audio classes in Audioset evaluation subset. The method used to build the descriptions is \textbf{Concat}, as explained in Sec.~\ref{ssec:3_1}}
	\label{fig:1}
	\vspace{-0.2cm}
\end{figure}

The inspiration for this work comes from the semantic similarity we observed in the label texts of Audioset. As shown in  Fig~\ref{fig:1}, we visualize the t-SNE projections of the text embeddings extracted from 527 audio classes in Audioset by a pretrained BERT \cite{devlin2018bert} model. To facilitate the observation, each projection is color-coded by its highest parent class according to Audioset ontology\footnote{https://research.google.com/audioset/ontology/index.html}. These parent classes are ``Human sounds'', ``Animal'', ``Music'', ``Source-ambiguous sounds'', ``Sounds of things'', ``Natural sounds'' and ``Channel, environment and background''.

In the t-SNE plot, different events naturally form distinct clusters based on their semantic similarities. After coloring each projected point according to the hierarchical relationships in the Audioset ontology, the projections still exhibit clustering characteristics, indicating that the semantic relationships of label text can describe the hierarchy and proximity relationships of the sound events. This finding highlights the potential of leveraging semantic information in audio tagging.


\section{Method}
\label{sec:Method}

\begin{figure}[ht]
	\centering
	\includegraphics[width=8.5cm]{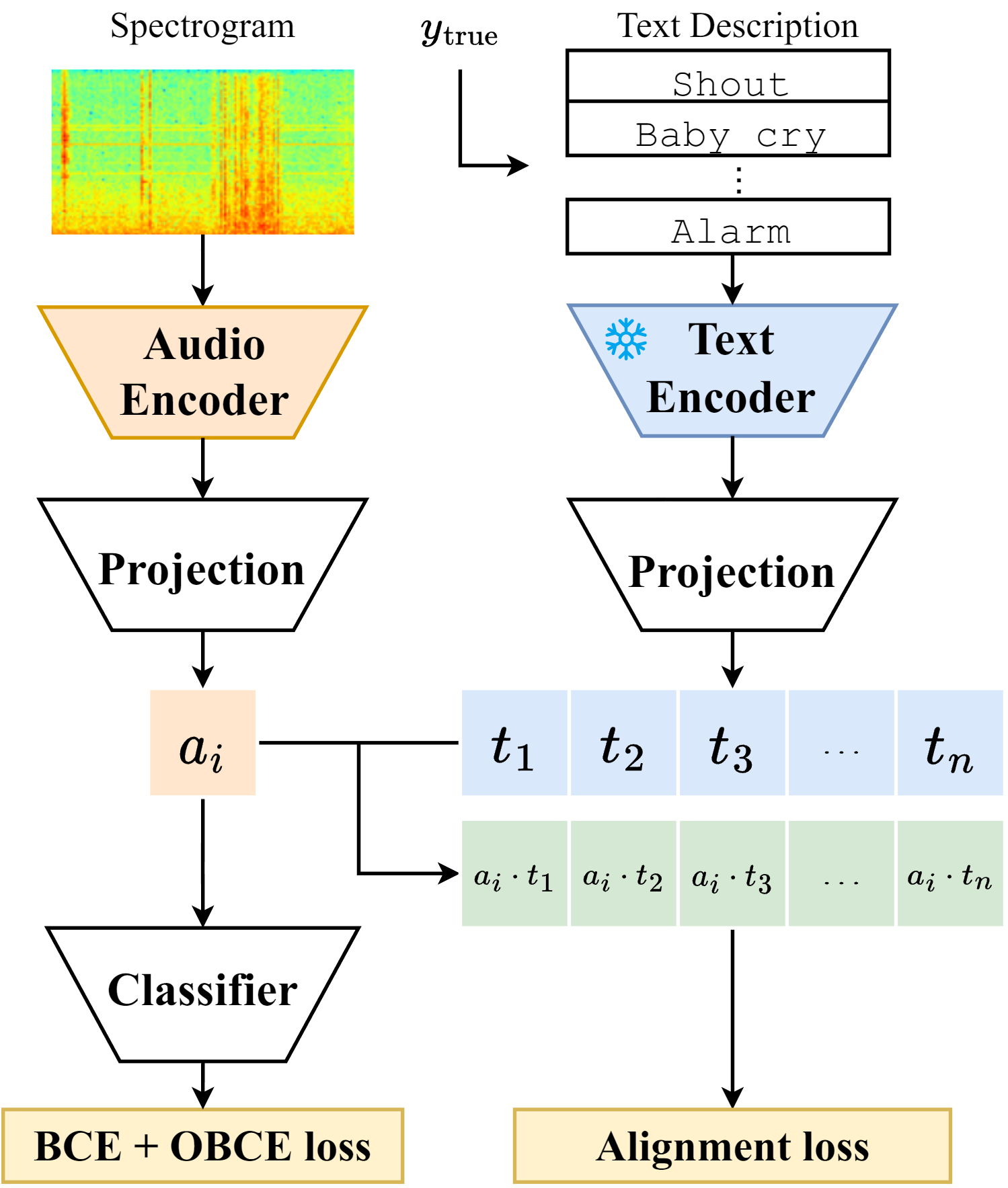}
	\caption{Architecture of proposed Semantic Proximity Alignment (SPA). We employ a pretrained BERT \cite{devlin2018bert} model as text encoder, which is kept frozen during training.}
	\label{fig:2}
	\vspace{-0.3cm}
\end{figure}

In order to align audio features with corresponding label text features, we use two pretrained encoders with projection layers to connect audio clips and language descriptions. Then we calculate the cosine similarity between audio feature and text features of its positve labels. The projected audio feature is then fed into a linear classifier to obtain the prediction over all target classes. The overall optimization objective is to minimize the classification losses while maximizing the similarity between audio features and corresponding text features. The structure of the method is shown in Figure~\ref{fig:2}.

\subsection{Building Label Description}
\label{ssec:3_1}
According to the available information in Audioset ontology, we employed four approaches to build the language description for each class:
\begin{itemize}[noitemsep]
    \item \textbf{Direct}: Use the display name of each class. (e.g. \texttt{\{label\}})
    \item \textbf{Prompt}: Use a combination of prompt and display name. (e.g. \texttt{the sound of \{label\}})
    \item \textbf{Desc}: Use the description of each class. (e.g. The description of \texttt{Male speech} is ``Speech uttered by an adult male human.'')
    \item \textbf{Concat}: For each class, choose the shortest path between the class and its highest parent class in the ontology graph. Concatenate all of its parent classes and itself in order. (e.g. For class \texttt{Male speech}, its language description will be ``Human sounds $>$ Human voice $>$ Speech $>$ Male speech''
\end{itemize}

\subsection{Formulation of SPA}
\label{ssec:2_1}
Given an audio clip $x$, we first extract the log-mel spectrogram $X$ s.t. $X \in \mathbb{R}^{F \times T}$ from $x$, where $T$ and $F$ refer to time frames and mel bins, respectively. Then, for $N$ positive labels of $x$, we select the descriptions of its positive labels from pre-defined label description map, forming the text sequence $L = [l_1, l_2, ..., l_N]$. Let $f_a(\cdot)$ be the audio encoder and $f_t(\cdot)$ be the text encoder, we can get the text representations $\hat{T}$ and audio representation $\hat{X}$:
\begin{equation}
    \hat{X}=f_a(X); \hat{T}=f_t(L)
\end{equation}
where $\hat{X} \in \mathbb{R}^{1 \times U}$ and $\hat{T} \in \mathbb{R}^{N \times V}$. $U$ and $V$ refer to the dimensions of the output representations from each encoder.
Two learnable linear projectors, $p_a$ and $p_t$, are then applied to project audio and label text representations into a joint space of dimension $D$ to get audio embedding $E_a$ and label text embeddings $E_t$:
\begin{equation}
    E_a=p_a(\hat{X}); E_t=p_t(\hat{T})
\end{equation}
where $E_a \in \mathbb{R}^{1 \times D}$ and $E_t \in \mathbb{R}^{N \times D}$.
The cosine similarity $S$ s.t. $S \in \mathbb{R}^{1 \times N}$ between the audio embedding and text embeddings is then calculated with $E_a$ and $E_t$.
The audio embedding is further fed into a linear classifier $c_a$ to obtain the prediction $\hat{\boldsymbol{y}} = c_a(E_a)$ over all target classes.

\subsection{Loss function}
\label{ssec:2_2}
The loss function consists of three components: Binary Cross Entropy (BCE) loss $\mathcal{L}_{\text {BCE}}$, Ontology-aware Binary Cross Entropy (OBCE) loss $\mathcal{L}_{\text {OBCE}}$, and alignment loss $\mathcal{L}_{\text {SPA}}$. 
$\mathcal{L}_{\text {OBCE}}$ proposed in \cite{liu2023ontology} reweights the loss for each class based on the ontology distance of different target labels with class weight $\boldsymbol{r}$, assigning a smaller weight to false predictions that are closer to the target labels. Given the target $\boldsymbol{y}$ and prediction $\hat{\boldsymbol{y}}$ of an audio clip, $\mathcal{L}_{\text {OBCE}}$ can be formulated as:
\begin{equation}
    \mathcal{L}_{\text {OBCE}}=\operatorname{mean}(\boldsymbol{r} \odot(\boldsymbol{y} \odot \log (\hat{\boldsymbol{y}})+(1-\boldsymbol{y}) \odot \log (1-\hat{\boldsymbol{y}})))
\end{equation}
To utilize the semantic information in text embeddings, we enforce cosine similarity for alignment loss $\mathcal{L}_{\text {SPA}}$ with audio embedding $E_a$ and label text embeddings $E_t$ as:
\begin{equation}
    \label{eq:4}
    \mathcal{L}_{\text {SPA}}=1-\frac{1}{N}\sum_{i=1}^N\frac{E_a \cdot E_{t_i}}{\left\|E_a\right\| \cdot\left\|E_{t_i}\right\|}
\end{equation}
$\mathcal{L}_{\text {SPA}}$ is assigned with a hyper-parameter $\alpha$. The final loss function $\mathcal{L}$ can therefore be formulated as:
\begin{equation}
    \label{eq:8}
    \mathcal{L}=(\mathcal{L}_{\text {BCE}} + \mathcal{L}_{\text {OBCE}})/2+\alpha\mathcal{L}_{\text {SPA}}
\end{equation}

\section{Experiments}
\label{sec:experiments}

\subsection{Training setup}
The model is trained on Audioset \cite{gemmeke2017audio}. All the audio clips are resampled into 32 kHz. For each audio clip, 128-dimensional log-mel spectrogram is extracted with a Hanning window. The window size is set to 800 with hop length set to 320. Other training details are the same with \cite{koutini2021efficient}.

Besides mAP, we utilize OmAP \cite{liu2023ontology} as evaluation metric. OmAP views the Audioset ontology as a complete graph, in which the distance between two nodes is set to be the smallest number of edges to connect them. OmAP is calculated at multiple coarse-grained levels $\lambda$, where $\lambda = 0, 1, ..., N - 1$. $N$ refers to the maximum distance between two arbitrary nodes in the ontology graph, which is $22$ for Audioset. $\text{OmAP}_\lambda$ will not take false positives (FP) into account if the distance between the FP and the target is smaller than or equal to $\lambda$. In the sections below, we use OmAP to refer to the average of all $\text{OmAP}_\lambda$.

We select five state-of-the-art audio tagging methods \cite{gong2021psla, kong2020panns, gong2021ast, koutini2021efficient, chen2022hts} as audio encoders. These audio encoders are all pretrained on Audioset. The MLP projector in text branch is jointly trained with the audio branch under the loss proposed in Eq.~\ref{eq:8}, while the text encoder is always kept frozen.

\begin{table}[ht]
  \centering
  \caption{mAP and OmAP (higher is better) of five selected models.}
  \label{tab:1}
  \begin{tabular}{@{}lcccc@{}}
    \toprule
    \makecell[c]{Audio\\Encoder} & \multicolumn{2}{c}{mAP (\%)}        & \multicolumn{2}{c}{OmAP (\%)} \\ \midrule
                                      & w/o SPA  & w/ SPA         & w/o SPA    & w/ SPA      \\ \cmidrule(){2-5}
    PSLA \cite{gong2021psla}          & 43.9   & 42.7 (-1.2)      & 76.2        & 76.9 (+0.7)     \\
    PANN \cite{kong2020panns}         & 43.3   & 42.4 (-0.9)      & 75.7        & 76.5 (+0.8)     \\
    AST  \cite{gong2021ast}           & 44.7   & 43.3 (-1.4)      & 77.3        & 78.8 (+1.5)     \\
    PaSST \cite{koutini2021efficient} & 46.8   & 45.7 (-1.1)      & 78.6        & 80.4 (+1.8)     \\
    HTS-AT \cite{chen2022hts}         & 45.9   & 44.6 (-1.3)      & 78.1        & 79.4 (+0.7)     \\ \bottomrule
  \end{tabular}
  \vspace{-0.5cm}
\end{table}

\subsection{Experimental Results}

\subsubsection{Objective performance}
\label{sec:experiments-1}

\begin{figure}[ht]
	\centering
	\includegraphics[width=8.5cm]{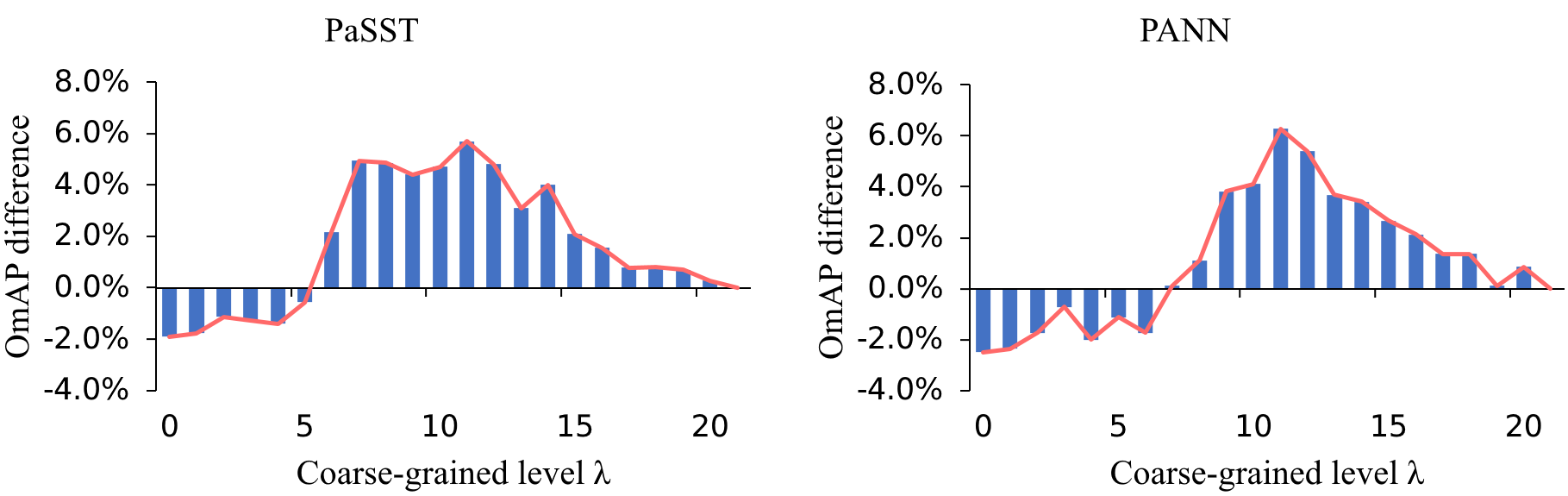}
	\caption{Differences of OmAP at each coarse-grained level $\lambda$ of PaSST \cite{koutini2021efficient} and PANN \cite{kong2020panns} with or without SPA.}
	\label{fig:3}
	\vspace{-0.5cm}
\end{figure}

As demenstrated in Tab.~\ref{tab:1}, the results of the experiment show that training with SPA has a positive impact on the OmAP scores of all the five SOTA models. This suggests that incorporating natural language information improves the models' ability to predict audio events at coarse-grained level. However, the drop in mAP seems inevitable despite multiple attempts of adjusting the weight $\alpha$ of $\mathcal{L}_{\text {SPA}}$. It is possibly because the text descriptions of each children classes share similar semantic information, while the more significant differences are primarily observed between parent classes. As a result, the text embeddings exhibit larger differences at a coarse-grained level and smaller differences at a fine-grained level, leading to an overall increase in OmAP but sacrificing the performance of fine-grained predictions. The analysis of $\text{OmAP}_\lambda$ at different coarse-grained levels $\lambda$ confirms this point, as shown in Fig.~\ref{fig:3}.

\subsubsection{Different language constructions}
\label{sec:experiments-2}

As described in Sec.~\ref{sec:2}, we adopt four different approaches to construct the language description of each class. As shown in Tab.~\ref{tab:2}, the best construction method of language description is to concatenate the label text of the target class and its parent classes. The explicit expression of hierarchical relationships among audio classes may help the model understand the hierarchical relationships between different label texts. Directly using label texts or descriptions from the Audioset ontology has positive effects on most models. However, using language descriptions constructed with prompts leads to a performance decline in all models. It is possible that for a fixed-weight text encoder, consistent prompts weaken the differences between text embeddings of different classes. This approach may yield better results if we jointly train the text encoder like \cite{wu2022wav2clip, guzhov2022audioclip, radford2021learning}.

\begin{table}[ht]
\centering
\caption{OmAP (\%) of models trained with four different approaches of constructing the language description of each class. }
\label{tab:2}
\begin{tabularx}{\columnwidth}{Xcccc}
\toprule
                                   & Direct          & Prompt        & Desc          & Concat         \\ \midrule
PSLA \cite{gong2021psla}           & 76.4            & 75.8          & 76.4          & \textbf{76.9}  \\
PANN \cite{kong2020panns}          & \textbf{76.5}   & 75.4          & 76.1          & 76.3           \\
AST  \cite{gong2021ast}            & 78.2            & 77.0          & 78.4          & \textbf{78.8}  \\
PaSST \cite{koutini2021efficient}  & 79.2            & 78.1          & 79.7          & \textbf{80.4}  \\
HTS-AT \cite{chen2022hts}          & 77.4            & 76.9          & 78.8          & \textbf{79.4}  \\ \bottomrule
\end{tabularx}
\vspace{-0.5cm}
\end{table}

\subsubsection{Human perception consistency}
\label{sec:experiments-3}

We observe certain differences in the predictions of the models with or without SPA. Therefore, We conducted a human perception experiment to evaluate which prediction aligns more closely with human perception. From the outputs of the models trained with or without SPA, we select the top-5 classes with the largest average differences on precision. For each class, we randomly sample 20 clips. Knowing the label of each sample, four participants are asked to decide whether the corresponding event is present in each sample, along with their confidence level (from 1 to 5). The opinions of the participants are then converted into human annotation $\boldsymbol{y}^{\text{human}}$, where presence is marked as 1 and absence as 0. The consistency score is defined as the average precision calculated with $\boldsymbol{y}^{\text{human}}$ and model outputs.

The selected top-5 classes with the highest average differences are: ``Air brake'', ``Child speech, kid speaking'', ``Fire'', ``Mechanisms'', ``Yodeling''. Among selected 100 clips, 89 of them have consistent annotations from at least three participants. The average confidence level is 4.355. Tab.~\ref{tab:4} presents the consistency scores of five models with or without SPA. 

\begin{table}[ht]
  \caption{Average consistency score (\%) of models with or without SPA. \textit{Audioset label} refers to $\hat{\boldsymbol{y}}$ generated based on whether the given event exists in its Audioset annotation for each sample, where presence is marked as 1 and absence as 0.}
  \label{tab:4}
  \centering
    \begin{tabularx}{\columnwidth}{Xcc}
    \toprule
                                   & without SPA        & with SPA   \\ \midrule
PSLA \cite{gong2021psla}           & 70.05           & 85.44   \\
PANN \cite{kong2020panns}          & 68.57           & 80.31   \\
AST  \cite{gong2021ast}            & 72.44           & 82.17   \\
PaSST \cite{koutini2021efficient}  & 71.56           & 83.52   \\
HTS-AT \cite{chen2022hts}          & 68.73           & 85.62   \\ \midrule
\textit{Audioset label}            & \multicolumn{2}{c}{61.19} \\ \bottomrule
    \end{tabularx}
\end{table}

As shown in Tab.~\ref{tab:4}, models show better consistency with human opinion after training with SPA. Additionally, we test the Audioset labels of each sample with human annotations $\boldsymbol{y}^{\text{human}}$. As mentioned in Sec.~\ref{sec:intro}, Audioset is plagued by a significant label noise issue. Out of the selected 100 samples, 71 samples were identified as having events by three or more annotators, while according to the Audioset labels, only 37 samples were labeled as having events, which leads to the surprising result in Tab.~\ref{tab:4}. This significant discrepancy further emphasizes the necessity of this study.

\section{Conclusion}

In this paper, we explore the impact of training audio tagging models with auxiliary text descriptions of sound events. Instead of directly mapping audio features to one-hot labels, we first extract the text features of pre-constructed language description of each audio class with a pretrained language model, and then align the audio feature of each clip with corresponding text features of its positive labels using proposed SPA loss. Experimental results demonstrate that aligning audio models with natural language leads to improved OmAP compared to the original models. Additional human perception experiments validate that the outputs of the models trained with SPA exhibit better alignment with human opinions. These results highlight the potential of leveraging natural language as auxiliary supervision to enhance the performance and perceptual quality of audio tagging models.


%
%
%



\let\oldthebibliography\thebibliography
\let\endoldthebibliography\endthebibliography
\renewenvironment{thebibliography}[1]{
  \begin{oldthebibliography}{#1}
    \setlength{\itemsep}{0.6em}
    \setlength{\parskip}{0em}
}
{
  \end{oldthebibliography}
}

\bibliographystyle{IEEEbib}
\bibliography{refs}

\end{document}